\documentclass[aps,prd,twocolumn,superscriptaddress]{revtex4}
\usepackage[dvips]{color,graphicx}

\begin{document}


\title{Atmospheric gamma-ray observation with the BETS detector
         for calibrating
  atmospheric neutrino flux calculations}


\author{K. Kasahara}
\email[]{kasahara@icrr.u-tokyo.ac.jp}
\homepage[]{http://eweb.n.kanagawa-u.ac.jp/~kasahara}
\author{E. Mochizuki}
\affiliation{Shibaura Institute of Technology, Saitama, Japan}

\author{S. Torii}
\author{T. Tamura}
\author{N. Tateyama}
\author{K. Yoshida}
\affiliation{Faculty of Engineering, Kanagawa  University, Yokohama, Japan}

\author{T. Yamagami}
\author{Y. Saito}
\author{J. Nishimura}
\affiliation{Institute of Space and Astronautical Science, Sagamihara, Japan}
\author{H. Murakami}
\affiliation{Department of Physics, Rikkyo University, Toshima-ku, Japan}
\author{T. Kobayashi}
\affiliation{Department of Physics, Aoyama Gakuin University, Setagaya-ku, Japan}
\author{Y. Komori}
\affiliation{Kanagawa Prefectural College, Kanagawa, Japan}
\author{M.Honda}
\author{T. Ohuchi}
\affiliation{Institute for Cosmic Ray Research, Univ. of Tokyo, Kashiwa, Japan}
\author{S. Midorikawa}
\affiliation{Information Department, Aomori University, Aomori,Japan}
\author{T. Yuda}
\affiliation{Solar-Terrestrial Environment Laboratory, Nagoya University,
        Aichi, Japan}


\date{\today}

\begin{abstract}
We  observed
atmospheric gamma-rays around 10 GeV
at balloon altitudes (15$\sim$25 km) and at a mountain (2770 m a.s.l).
 The observed results were compared with  Monte Carlo calculations 
 to find that an interaction model (Lund Fritiof1.6)
 used in an old neutrino flux calculation
 was not good enough for
 describing the observed values.  In stead, we found that two other
 nuclear interaction models, 
 Lund Fritiof7.02 and dpmjet3.03, gave much better agreement with the     
 observations.
 Our data will serve for examining nuclear interaction models
 and  for deriving a reliable
 absolute atmospheric neutrino flux in the GeV region.
\end{abstract}


\maketitle

\section{Introduction}

The discovery of evidence for neutrino oscillation by the Super Kamiokande 
group\cite{skoscillation} is based on the comparison of the observed atmospheric 
neutrino flux
with  calculated values.  Although the conclusion is so derived
that it would not be 
upset by the uncertainty of the absolute
flux value, it is desirable to obtain a reliable expected
neutrino flux (under no oscillation assumption) for further detailed discussions.

Two major sources of uncertainty in the atmospheric neutrino flux calculation
are 
1) the primary cosmic-ray spectrum and 2) the propagation
of cosmic rays in the atmosphere, especially, 
modeling of the nuclear interaction. The  absolute flux calculations so far
made
by various groups are expected to have  uncertainty of  $\sim$ 30 \%\cite{GHreview}.

The primary proton and  He spectra recently measured with magnet
spectrometers  by the
 BESS
\cite{bess1ry}
and AMS\cite{ams1ry} 
groups agree  very well and seem reliable.   Therefore, we
may take that the first
problem mentioned above have now been almost settled at least up to 100 GeV/n.
This means that if we have a reliable atmospheric cosmic-ray flux data,
we may compare it with a calculation which uses such primaries and
test the validity of nuclear interaction models.

For such an atmospheric cosmic-ray component, one may first raise the muon
and actually some new observations 
have been or being tried\cite{capricemuon1, capricemuon2, bessmuonnori}.

As a secondary cosmic-ray component, we focused 
on gamma-rays which are easy to measure with
our detector.  A good model should be able to explain
muons  and gamma-rays simultaneously.  Muons are important
since they are directly
coupled with neutrinos, but the flux is affected somehow by
the structure of the atmosphere 
 which is usually
not well known.  Compared to muons, the flux of gamma-rays is substantially
lower but
is almost
insensitive to the atmospheric structure and depends only on
the total thickness to the  observation height.

In 1998, we performed first gamma-ray observation with our
detector
 at Mt. Norikura
(2770m a.s.l) in Japan, and also made subsequent two successful observations
at balloon altitudes (15 $\sim 25$ km) in 1999 and 2000. 
In the present paper, we report the final results of 
these observations and consequences.

\section{The Detector}
For our observation, we upgraded the BETS
(Balloon-born Electron Telescope with Scintillating fibers)
detector which had
been developed for the observation of cosmic primary electrons
in the 10 GeV region.
Its details before being upgraded for gamma-ray observation is
in \cite{betsnim} and the electron observation result is in
\cite{betselec}.
The basic performance  was tested at CERN using
electron, proton and pion beams of 10 to 200 GeV\cite{betsnim, betscern}.
Although this was undertaken before the upgrading, 
we can essentially use that calibration for the
current observeions partly with a help of Monte Carlo simulations.

\begin{figure}[h]
 \includegraphics[width=92mm]{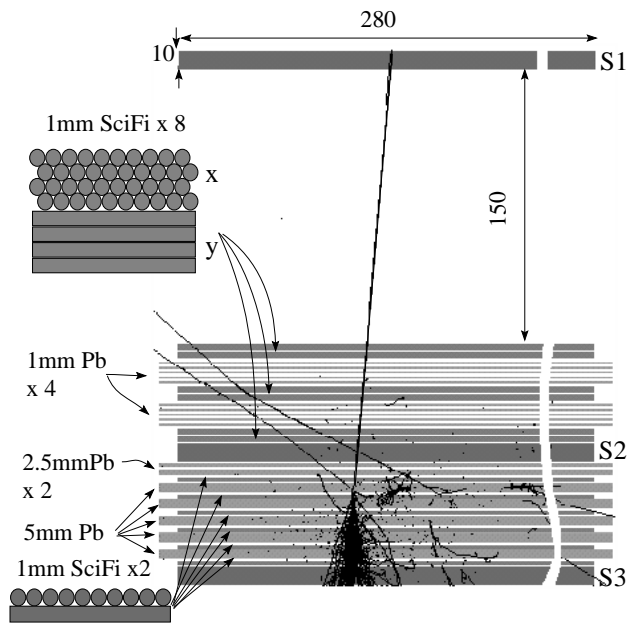}
 \caption{Schematic illustration of the main body of the detector.
S1, S2 and S3 are 1 cm thick plastic scintillators used for
trigger. Each fiber has 1mm diameter.
Originally nuclear emulsion plates were placed on the upper scifi's
and also inserted between the upper
        thin lead plates for detailed investigation of
        tracking capability of scifi.  They are kept
        in the present system to have the same structure at
        the calibration time.  The inlaid cascade shows charged
        particle tracks  by a simulation for a 30 GeV incident  proton.
\label{det}}
 \end{figure}

Figure \ref{det} shows a schematic structure of the main body of BETS.
The
calorimeter has
7.1 r.l lead thickness and the cross-section
is 28 cm $\times$ 28 cm. The whole detector system is 
contained in a pressure vessel made of thin aluminum.

\begin{table}[h] 
\begin{center}
\caption{Basic characteristics of BETS\\
        (triple numbers in the table are for gamma-ray 
        energy of 5, 10, and 30 GeV, respectively)
\label{basicchara}}
\begin{tabular}{|c|c|}
\hline
 R.M.S energy resolution(\%) &  21, 18, 15 (for $\theta\sim 15^\circ$) \\
\hline
 S$\Omega$(cm$^2$sr) & 243, 240,218 (at $\sim$20 km)  \\
\hline
 R.M.S angular resolution (deg) & 2.3, 1.3, 1.0 (for $\theta\sim 15^\circ$) \\
\hline
 Total number of  scifi's & 10080  \\
\hline
 Weght including electronics (kg) & 230  \\
\hline
 Cross-section of the main body & 28cm $\times $ 28cm  \\
\hline
 Thickness (Pb radiation length) & 7.1  \\
\hline
\end{tabular}
\end{center}
\end{table}

 The main feature of the BETS detector is that it is
 a tracking calorimeter; it contains a number of sheets consisting of
 1 mm diameter scintillating fibers (scifi), many of which are
 sandwiched between lead plates.  
 The total number of scifi's
 are 10080.  The sheets are grouped into two types; 
 one is  to serve for x and  the other for  y position
 measurement.  Each of them is fed to 
 an image intensifier which in turn is connected to a CCD.  Thus, 
 the two CCD output gives us an $x-y$ image of cascade shower development
 and  enables us to 
 discriminate gamma-rays, electrons from
 other (mainly hadronic) background  showers.
 The proton rejection power against electron is
 $R\sim 2\times 10^3$ (i.e,
 one misidentification among $R$ protons)       
at 10 GeV\footnote{%
We note electron showers of 10 GeV are
normally simulated by $\sim$ 30 GeV protons
when the latter start cascade at a shallow depth of 
the detector.}
 The basic characteristics of the detector
       are summarized in Table \ref{basicchara}.
%

 \begin{figure}[h]
 \includegraphics[width=85mm]{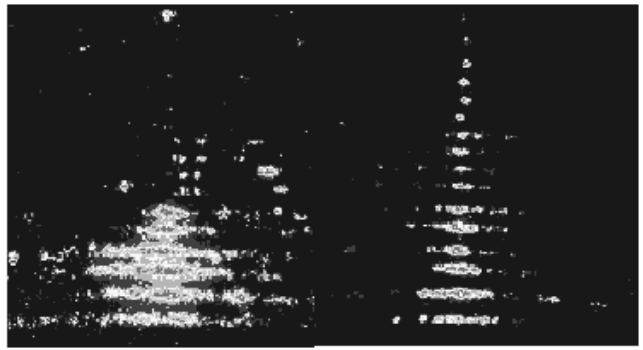}%
 \caption{Image of cascade shower by a proton (120 GeV,left)
        and an electron(10 GeV, right) obtained at CERN.
\label{image}}
 \end{figure}

 In Fig.\ref{image}, we show  examples of the CCD image of  a
 cascade shower for
 a proton incident case and  for an electron
 incident case.

\begin{figure}[h]
 \includegraphics[width=85mm]{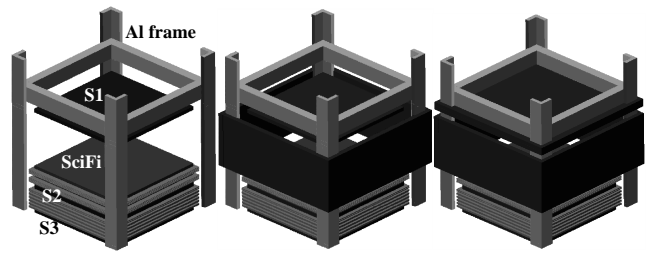}
 \caption{Yearly change of the anti-counters.
 Left: 1998. No change from original BETS except for
        trigger logic. Middle: 1999. 1.5 cm thick plastic scintillator
        side anti-counters  were added.  Right: 2000.
        The whole top view was covered by a 1 cm thick plastic
        sintillator.
\label{anti}}
 \end{figure}

Figure \ref{anti}  illustrates the yearly change of anti-counters.
In 1998 (Mt.Norikura observation), the main change was limited to
the upgrading of trigger logic.   In 1999, we
added 4 side anti-counters (each 15 cm $\times$ 36 cm $\times$
 1.5 cm plastic scintillator.
 Nine optical fibers containing wave length shifter
are embedded in each scintillator and  connected  to a Hamamatu H6780
PMT. 

 In 2000, we further
added an anti-counter which covers the whole  top view of the
detector and also improved data acquisition speed.
The top anti-counter is 38 cm $\times$ 38 cm $\times$ 1 cm plastic
scintillator. We also embedded optical fibers; 8 in the $x$ and another
8 in the $y$ direction, all of which were fed to an H6780.

Although we could remove  background showers without the anti-counters,
inclined particles (mainly protons) entering from the gap
between top  scintillator (S1) and the main body
degrades  the desired gamma-ray event rate. 
The addition of the top anti-counter greatly helped improve
this rate.

We emphasize that detection of gamma-rays is easier for
us than that of electrons, since, for  gamma-rays,
we can utilize  absence of incident charge.

\section{Observations}

Table \ref{sumtab} shows the summary of the observations.

\begin{table*}[t] 
\begin{center}
\caption{Summary of three observations\label{sumtab}}
\begin{ruledtabular}
\begin{tabular}{|l|c|c|c|c|c|c|c|c|c|c|}
\hline
Observation &  Mt.Norikura(1998) & \multicolumn{5}{c|}{Balloon (Sanriku, 1999)} & 
                                \multicolumn{4}{c|}{Balloon (Sanriku, 2000)} \\
\hline
Period & Aug.31$\sim$Sep.18 & \multicolumn{5}{c|}{Sep.2, 6:55$\sim$17:17} & 
                \multicolumn{4}{c|}{Jun.5, 6:30$\sim$17:59} \\
\hline
Altitude(km) & 2.77 &15.3 & 18.5 & 21.2 & 24.7 & 32.3 & 15.3 & 18.3 & 21.4 & 25.1 \\
\hline
Depth(g/cm$^2$) &  737 & 126 & 74.8 & 48.9 & 28.0 & 9.5 & 128 & 73 & 45.7 & 25.3 \\
\hline
Obs. hour (s) & $1.33\times 10^6$ & 1260 & 1560 & 2100 & 4878 & 3120 &
        1560 & 2160 & 4320 & 2320 \\
\hline
Live time (s) & $9.8\times 10^5$ & 504 & 450 & 414 & 852 & 498 &
                752 & 928 & 1805 & 789 \\
\hline
Live time (\%) & 74.0 & 40.0 & 28.8 & 19.7 & 17.5 & 16.0 & 
                48.2 & 43.0 & 42.6 & 44.2 \\
\hline
Triggered events & $1.8\times 10^6 $ & 9513 & 11288 & 13361 & 30439 & 16741 &
                18808 & 25795 & 46675 & 17436 \\
\hline
$\gamma$ events & $4.7\times 10^4$ & 700 & 650 & 611 & 848 & 345  &
                1300 & 1485 & 2299 & 740 \\
(\%)    & 2.5  & 7.3    &5.7 & 4.6 &  2.8 & 2.0 &
                        6.9 & 5.8 & 4.9 & 4.2  \\
\hline
g-low trigger     & S1 $< 0.5$ & \multicolumn{5}{c|}{S1$<0.5$} &
                      \multicolumn{4}{c|}{S1$<0.47$} \\
condition (in mip).        & S2 $> 2.3$ & \multicolumn{5}{c|}{S2$>1.5$} &
                         \multicolumn{4}{c|}{S2$>1.59$} \\
                  & S3 $> 1.7$ & \multicolumn{5}{c|}{S3$>3.0$} &              
                         \multicolumn{4}{c|}{S3$>3.18$} \\
\hline
\end{tabular}
\end{ruledtabular}
\end{center}
\end{table*}

\begin{figure}[h]
 \includegraphics[width=73mm]{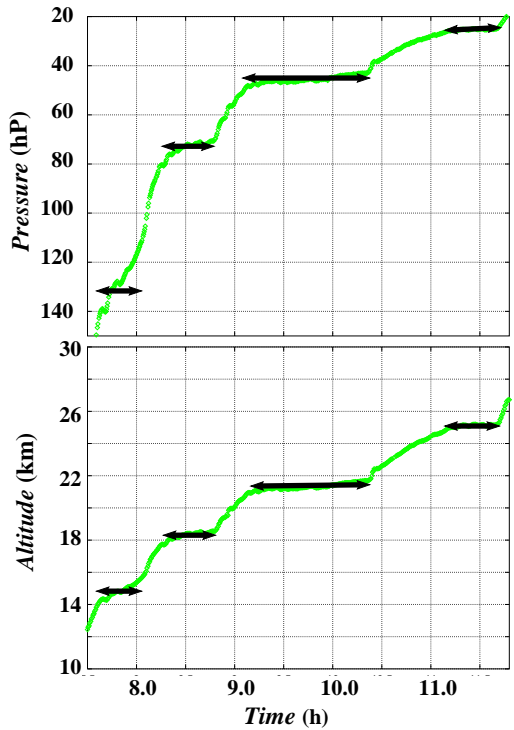}
 \caption{Flight curve of the 2000 observation. 
        Pressure (upper) and  altitude (lower) as a function
        of time. Each arrow  shows the level flight region.
        The pressure change at around 15.3 km is rather
        rapid but the gamma-ray intensity is almost constant there and
        the change can be neglected. 
\label{flight}}
 \end{figure}

\begin{itemize}
\item Mt. Norikura observation.

        Our first  gamma-ray observation was
        performed in 1998 at Mt.Norikura Observatory of Univ. of
        Tokyo, Japan (2770 m a.s.l, latitude 36.1$^\circ$N, 
        longitude 137.55$^\circ$E, magnetic cutoff rigidity  $\sim$ 11.5 GV).
         The atmospheric pressure during
        the observation is shown in Fig.\ref{noripress}.
        The average atmospheric depth is 737 g/cm$^2$.

\begin{figure}[h]
 \includegraphics[width=85mm]{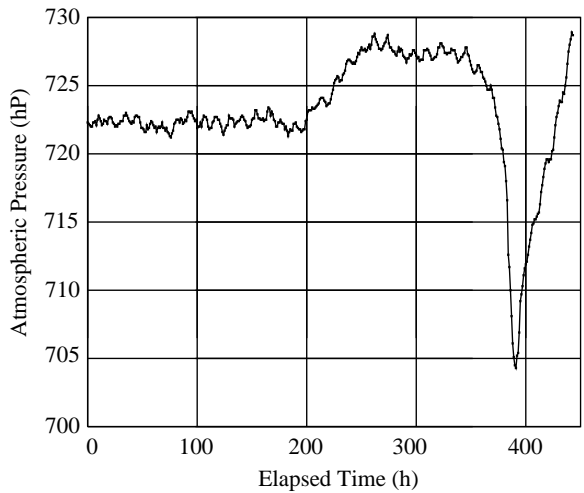}
 \caption{Pressure change during Mt. Norikura observation.
 The last pressure drop is due to a typhoon.
 The average pressure is 723 hP (737 g/cm$^2$).
\label{noripress}}
 \end{figure}

\item Balloon flight

        We had two similar balloon filights in 1999 and 2000.
        Since the main outcome of the  data is  from
        the latter, we briefly describe it.
        A balloon of  43$\times 10^3$ m$^3$
        was launched at 6:30 am, 5th June, 2000
        from the Sanriku balloon center of the
        Institute of Space and Astronautical Science,
        Japan (latitude 39.2$^\circ$N, longitude
               141.8$^\circ$E, magnetic cutoff rigidity $\sim$ 8.9 GV)
         and recovered  with the help of the helicopter.
         at
        17:59 on the
        sea  not far from the center.
        The flight curve shown in
        Fig.\ref{flight}  confirms that
        we have  good level flights
        at 4 different heights.

        As compared to the 1999 flight,
        this flight realized a smaller  dead time and higher ratio of
        desired gamma-ray events.
\end{itemize}

\subsection{Event trigger}
The basic event trigger condition is created by
signals from the three
plastic scintillators (S1, S2 and S3).  We show the discrimination
level in terms of the minimum ionizing particle number which 
is defined by the peak of the energy loss distribution of
cosmic-ray muons passing both S1 and S3 with inclination
less than 30 degrees.

We prepare a multi-trigger system by which event trigger
with different conditions is possible at the same time.
The major two trigger modes are the g-low  and g-high.
The g-low  is
responsible for low energy gamma-rays and
all anti-counters, when available, are used as veto counters.
Its condition is listed in Table \ref{sumtab}.
High energy gamma-rays  normally
produce a lot of back splash particles which hit S1 and/or
anti-counters, and thus the g-low trigger is suppressed.
In such a case, i.e, if we have a large S3 signal, anti-counter
veto is invalidated and the S1 threshold is relaxed
(The g-high condition is S1$<3.0$, S2$>5.0$ and S3$>8.1$).

The branch even point of the g-low and g-high mode efficiency
is at $\sim $30 GeV.  
Since we deal with gamma-rays mostly below 30 GeV, and also to avoid
complexity,  we present results only by the g-low mode.

\section{Analysis}

\subsection{Event selection}

Among the triggered events, we selected gamma-ray candidates 
by imposing the following conditions:

\begin{figure}[hbt]
\begin{center}
\vspace*{2.0mm} 
\includegraphics[width=8.5cm]{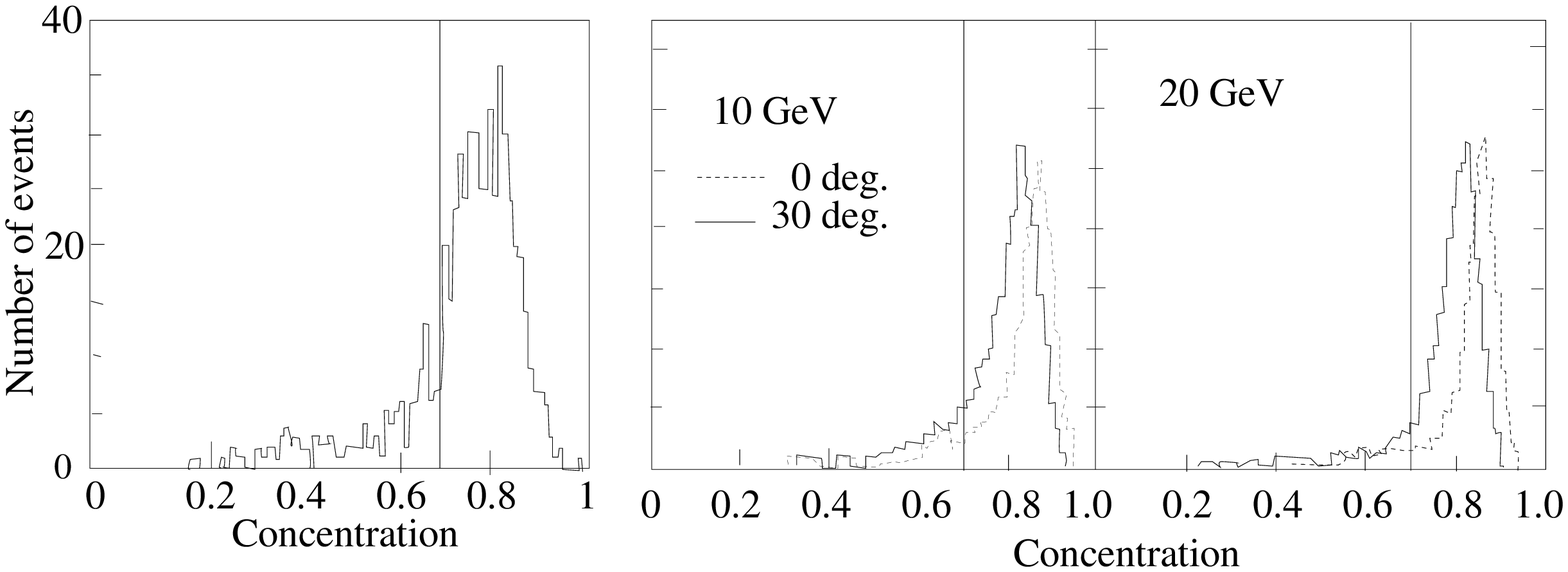} 
\caption{(left)Energy concentration distribution at 21.4 km.
 (right)the same by electrons at CERN }\label{conc}
\end{center}
\end{figure}

\begin{enumerate}
\item The estimated shower axis passes  S1 and S3. 
        The axis position in S3 must be at least 2 cm apart from
        the edge of S3.
\item The estimated shower axis has a zenith angle less than 30 degrees.
\item The energy concentration (see below) must be greater than 0.7.
\end{enumerate}

According to a simulation, only neutrons could be a background
against gamma-rays and the 3rd conditions above reduces  the
neutron contribution to a negligible level ($<1$\%).

The energy concentration is defined as the fraction of
scintillating fiber light intensity within 5 mm from the shower axis. Figure
\ref{conc} shows the concentration of analysed events together
 with the result of CERN data. Hadrons make a distribution with
 a peak at around 0.5. We see that the contribution of hadrons
in our observation is negligible.

\subsection{Energy Determination}
The energy calibration was performed in 1996 at CERN using 
electrons with energy 10 $\sim $ 200 GeV\cite{betsnim,betscern}.
There is no
direct calibration for gamma-rays, but, for the present
detector  thickness and energy range,
a M.C simulation
tells us that the calibration in 1996 can be used  for 
gamma-rays, too\footnote{%
        If we don't impose the trigger condition, the gamma-ray
        case shows a small difference from the electron case.
        }.
Therefore,
for the 1998 and 1999 observations,  energy is
obtained as a function of the S3 output and
zenith angle  using 
the CERN calibration.

\begin{figure}[h]
\includegraphics[width=8cm]{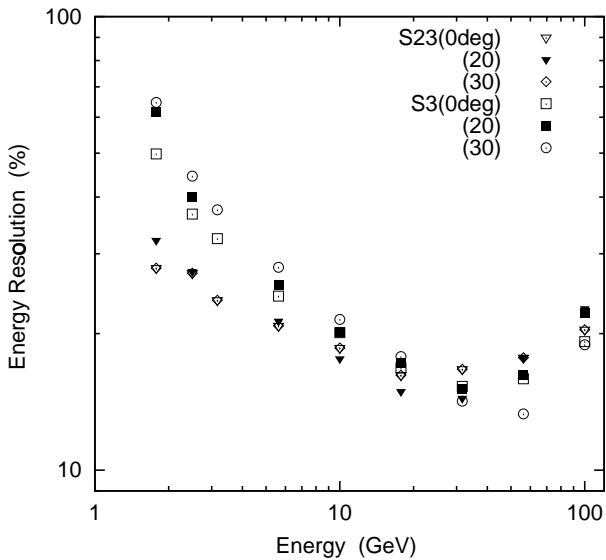} 
\caption{R.m.s energy resolution. 
The resolution by 
S2+S3 or S3 only is shown.
Different symbols indicate different incident
angles. We used S2+S3 below 20 GeV for the year 2000 
        data.
\label{eresol}}
\end{figure}

In 2000, we made some change in the
electronics so the CERN calibration
could not be used directly.  The effect by the change was
absorbed by a M.C simulation of which the validity
was verified by examining the 1998 and 1999 data.
We used
the sum of S2 and S3 outputs below 20 GeV since the
energy resolution was found to be better than using S3 only.
Figure \ref{eresol} shows
r.m.s energy resolution.

\subsection{Correction of the gamma-ray intensity}

The gamma-ray vertical flux is obtained from  the raw $dN/dE$
by  dividing it by the live time of the detector and
the effective $S\Omega$ (area $\times$ solid angle).
The latter is obtained by a simulation\cite{someganu00}.
It is dependent on the observation hight and energy.
A typical value at 10 GeV is 240 cm$^2$sr (see Table\ref{basicchara}).
The energy spectrum is further corrected by the
following factors which are not taken into account
in the $S\Omega$ calculation.

\begin{figure}[htb]
\begin{center}
\includegraphics[width=7cm]{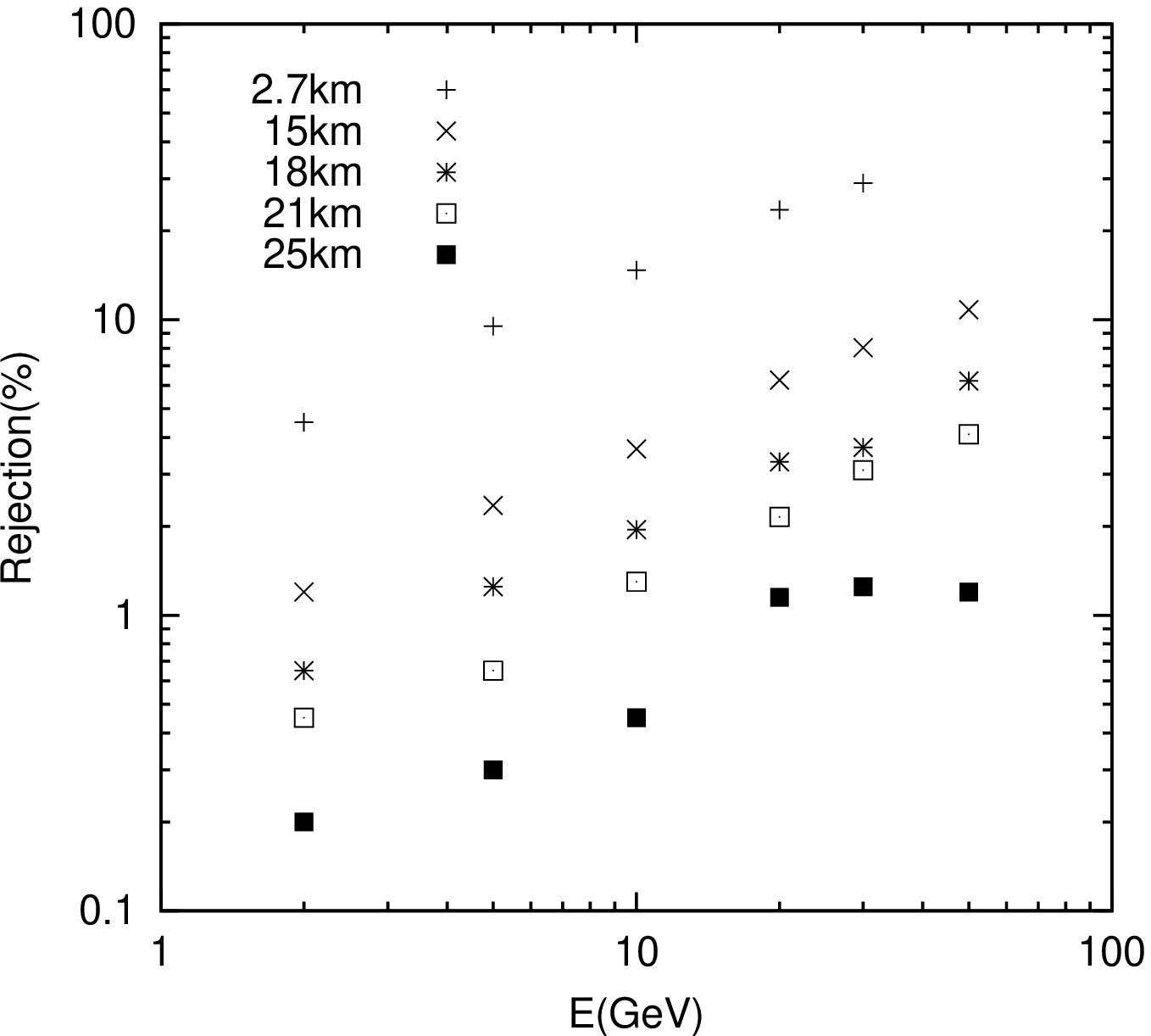} 
\includegraphics[width=7cm]{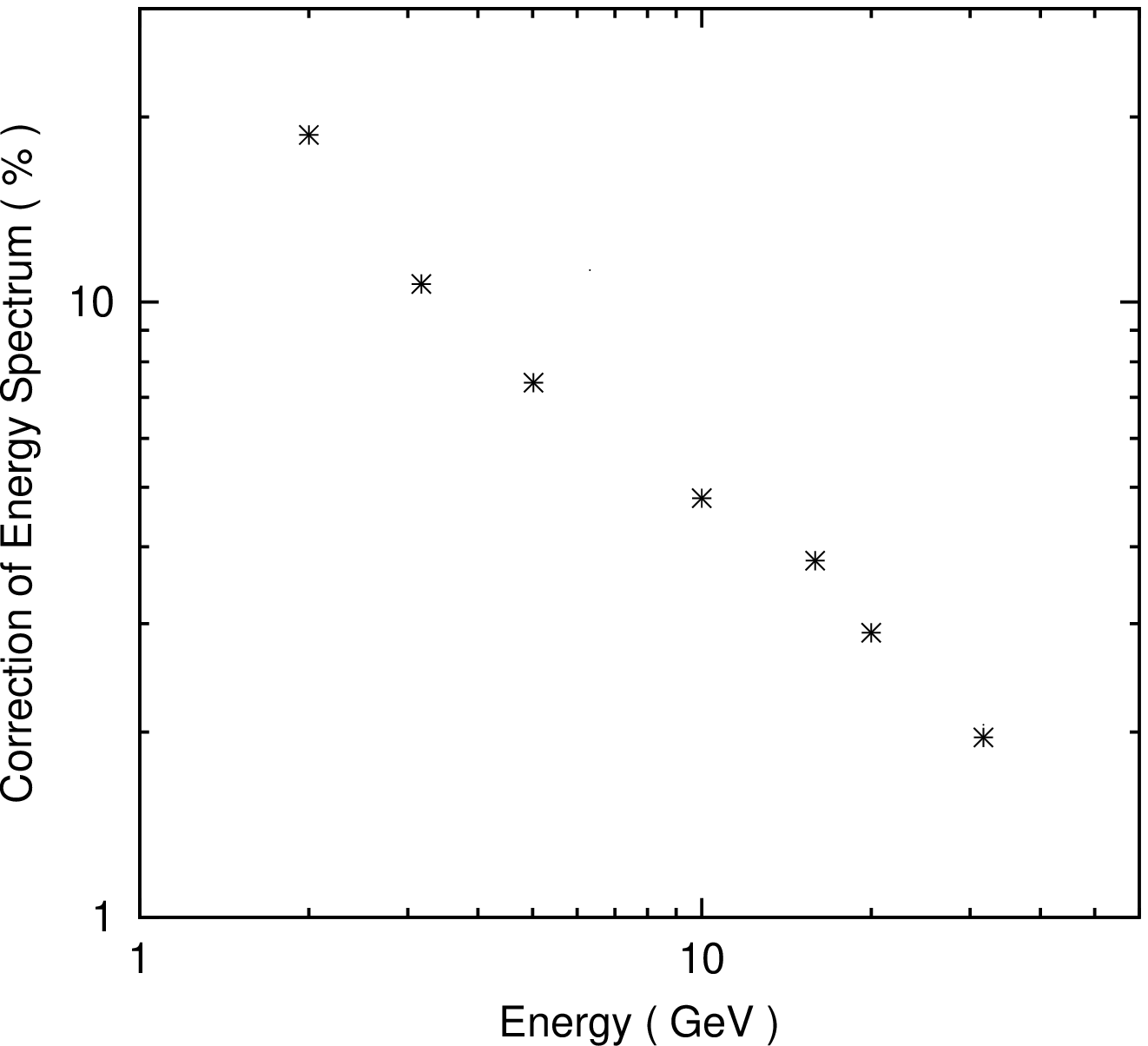} 
\caption{(upper)Multiple incidence rate. (lower)
  Correction factor for year 2000 due to spillover.
        The flux must be lowered. For Norikura,
        the factor below 20 GeV is larger by
        1$\sim 3$ \%.
        }\label{correc}
\end{center}
\end{figure}

\begin{enumerate} 
\setlength{\itemsep}{-0.15cm}
\item  Systematic bias in our estimation of the
        shower axis.  We underestimate the
        zenith angle systematically and it
        leads to overestimation of the intensity about 4\% 
        for the balloon and 1.8 \% for Mt.Norikura
        observations.

\item   Multiple incidence of particles.  A gamma-ray
        is sometimes accompanied by other charged particles
        and they enter the detector simultaneously
        (within 1 ns time difference in 99.9 \% cases).
        They are a family of particles generated by one
        and the same primary particle\footnote{%
                The chance coincidence probability of
                uncorrelated particles is 
                negligibly small.}.
         The charged particles fire the anti-counter
        and the g-low trigger is inhibited.
        
        In some case, multiple gamma-rays enter
        the detector simultaneously.  The rate is
        smaller than the  charged particle case.        
         However, this  is judged as a hadronic shower
        in most of cases.  The multiple incidence
        leads to the underestimation  of gamma-ray intensity.
        The portion of multiple incidence is
          shown in Fig.\ref{correc} (upper).

\item Finite energy resolution. The rapidly falling energy spectrum 
        leads to the spillover effect.
        This normally leads to the overestimation
        of flux (Fig.\ref{correc}, lower).
\end{enumerate}

\section{Results and comparison with calculations}
The flux values are summarized in Table \ref{flux}. 
We put only the statistical errors in the flux values, since
systematic
errors coming from the uncertainty of the S$\Omega$ calculation,
various cuts  and
flux corrections are expected to be order of a few percent
and much smaller than the present statistical errors.

\begin{table*}[t] 
\begin{center}
\caption{Summary of flux values\label{flux}}
\begin{ruledtabular}
\begin{tabular}{|l|l|l|l|l|l|l|l|l|l|}
\hline           
     \multicolumn{10}{|c|}{height (km)} \\
\hline
     \multicolumn{2}{|c|}{15.3} &    \multicolumn{2}{c|}{18.3} &    \multicolumn{2}{c|}{21.4} & 
   \multicolumn{2}{c|}{25.1} &    \multicolumn{2}{c|}{32.3}  \\
\hline
  \multicolumn{10}{|c|}{Energy (GeV)\  \&\  flux(No./$m^2\cdot$ s$\cdot$sr$\cdot$GeV)}   \\
\hline
5.48 & 2.42 $\pm$ 0.37  & 5.48 & 2.11 $\pm$ 0.39   & 5.47 & 2.11 $\pm$ 0.24  & 5.47  & 1.58 $\pm$ 0.25 & 5.47  & 0.49 $\pm$ 0.14 \\
6.47 & 1.18 $\pm$ 0.27  & 6.47 & 1.10 $\pm$ 0.24   & 6.47 & 1.35 $\pm$ 0.21  & 6.47  & 0.82 $\pm$ 0.18 & 6.57  & 0.19 $\pm$ 0.09 \\
7.47 & 0.89 $\pm$ 0.24  & 7.47 & 0.79 $\pm$ 0.21   & 7.47 & 0.82 $\pm$ 0.16  & 7.47 & 0.66 $\pm$ 0.16  & 7.47  & 0.24 $\pm$ 0.10 \\
8.48 & 0.37 $\pm$ 0.15  & 8.48 & 0.92 $\pm$ 0.20   & 8.48 & 0.51 $\pm$ 0.13  & 8.48 & 0.49 $\pm$ 0.14  & 8.48  &  0.16 $\pm$ 0.08 \\
9.48 & 0.54 $\pm$ 0.17  & 9.85 & 0.46 $\pm$ 0.11   & 9.48 & 0.50 $\pm$ 0.12  & 9.48 & 0.36 $\pm$ 0.12  & 9.48  &  0.16 $\pm$ 0.08 \\
10.5 & 0.17 $\pm$ 0.10  & 11.5 & 0.35 $\pm$ 0.12   & 10.5 & 0.41 $\pm$ 0.09  & 10.5 & 0.34 $\pm$ 0.12  & 12.3  & 0.13 $\pm$ 0.037 \\
12.1 & 0.28 $\pm$ 0.09  & 14.0 & 0.24 $\pm$ 0.06   & 11.8 & 0.23 $\pm$ 0.069 & 12.2 & 0.21 $\pm$ 0.054 & 17.0 & 0.032 $\pm$ 0.018 \\            
14.0 & 0.17 $\pm$ 0.05  & 18.3 & 0.072 $\pm$ 0.030 & 14.0 & 0.16 $\pm$ 0.030 & 14.0 & 0.076 $\pm$ 0.03 & 21.7 & 0.022$\pm$ 0.015 \\            
18.5 & 0.12 $\pm$ 0.04  & 26.8 & 0.040 $\pm$ 0.017 & 18.4 & 0.086 $\pm$ 0.023& 17.8 & 0.078 $\pm$ 0.029 &      &                  \\
25.5 & 0.06 $\pm$ 0.02  &      &                   & 27.1 & 0.026 $\pm$ 0.009& 21.7 & 0.064 $\pm$ 0.026 &      &                  \\
     &                  &      &                   &      &                  & 26.8 & 0.024 $\pm$ 0.012 &      &                  \\
     &                  &      &                   &      &                  & 36.0 & 0.012 $\pm$ 0.008 &      &                  \\
\hline
\end{tabular}
\end{ruledtabular}
\end{center}
\end{table*}

\begin{table}[h] 
\begin{center}
\caption{Flux values at Mt. Norikura\label{noriflux}}
\begin{tabular}{|c|c|}
\hline
E(GeV) & Flux ($10^{-4}/$m$^2\cdot$s$\cdot$sr$\cdot$GeV) \\
\hline
5.48 & 274 $\pm$ 13 \\
6.47 & 183 $\pm$ 11 \\
7.47 & 133 $\pm$ 9  \\
8.47 &  87.8 $\pm$ 7.5  \\
9.47 &  86.5 $\pm$ 7.5  \\
10.5 &  54.1 $\pm$ 5.9 \\
11.5 &  46.6 $\pm$ 5.5 \\ 
12.5 &  38.3 $\pm$ 5.0 \\ 
13.5 &  32.6 $\pm$ 4.6 \\ 
14.5 &  24.2 $\pm$ 4.0 \\ 
15.5 &  25.7 $\pm$ 4.1 \\ 
17.0 &  11.9 $\pm$ 2.0 \\ 
19.0 &  15.3 $\pm$ 2.3 \\ 
21.0 &  13.1 $\pm$ 2.1 \\ 
23.0 &   5.80 $\pm$ 1.4 \\ 
26.0 &   5.31 $\pm$ 0.95 \\ 
30.0 &   3.00 $\pm$ 0.72 \\ 
34.0 &   2.30 $\pm$ 0.64 \\ 
38.0 &   1.07 $\pm$ 0.44 \\ 
45.0 &   1.45 $\pm$ 0.32 \\ 
55.0 &   0.52 $\pm$ 0.20 \\ 
65.0 &   0.22 $\pm$ 0.13 \\ 
75.0 &   0.30 $\pm$ 0.15 \\ 
85.0 &   0.15 $\pm$ 0.10 \\ 
\hline
\end{tabular}
\end{center}
\end{table}

\begin{figure}[h]
\begin{center}
\includegraphics[width=6.5cm]{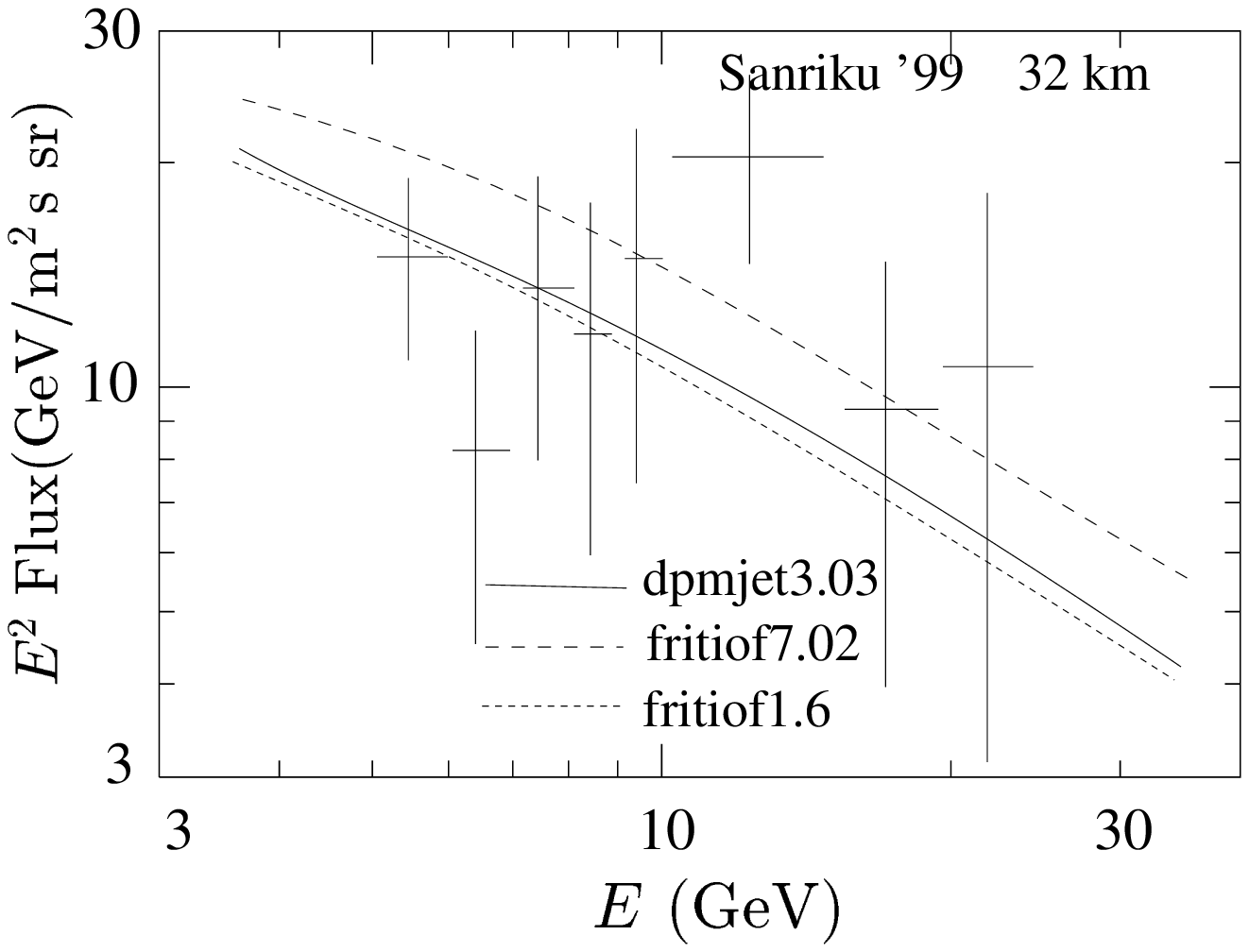}
\caption{Gamma-ray spectra at 5 balloon heights
are compared with 3 different models. The
vertical axis is Flux$\times E^2$.
 Except for
1999 data at 32.3 km,  1999 and 2000 flights data
are combined.
From
top to bottom, at 25.1, 21.4, 18.3, 15.3 and 32.3 km.
The spectra expected from three interaction models are
drawn by solid (dpmjet3.03), dash (fritiof7.02) and
dotted (fritiof1.6) lines. 
}
\label{balspec}
\end{center}
\end{figure}

The gamma-ray energy spectra thus obtained  at balloon altitudes
are shown in Fig.\ref{balspec} together with the expected
ones calculated by the Cosmos simulation code\cite{cosmos}.  
Except for 32.3 km altitude, we can disregard the small difference of
the observation depths and  we combine two flight data
with statistical weight, although the main contribution is
from the  flight in 2000.

  In the simulation calculation,
  we employed 3 different nuclear interaction models: 1) fritiof1.6\cite{oldfri}\footnote
        {It is used at energies greater than 4.5 GeV. At lower
        energies, nucrin/hadrin\cite{nucrin}
        is used.}
        used in the HKKM calculation\cite{hkkm95},
        which was widely used for comparison with the Kamioka data,
         2)fritiof7.02\cite{newfri}\footnote{
                It is used at energies greater than 10 GeV.
                At lower energies, model is the same as fritiof1.6}
 and 3) dpmjet3.03\cite{dpmjet}. As the primary cosmic ray,
        we used the  BESS result on protons and He. 
        The CNO component is also considered\cite{cno}.
         Besides these
        we included electron and positron data by AMS\cite{amselec}.
        Their data in the 10 GeV region is consistent
        with the HEAT\cite{heat} and BETS\cite{betselec} data.
        Bremstrahlung gamma-rays from the primary electrons could contribute
        order of $\sim 10$ \% at very high altitudes.

At balloon altitudes, the two models, fritiof7.02 and
dpmjet3.03, give almost the same results which are
        close to the observed data, while fritof1.6
 gives clearly smaller 
fluxes than the observation.

\begin{figure}[h]
\begin{center}
\includegraphics[width=7.5cm]{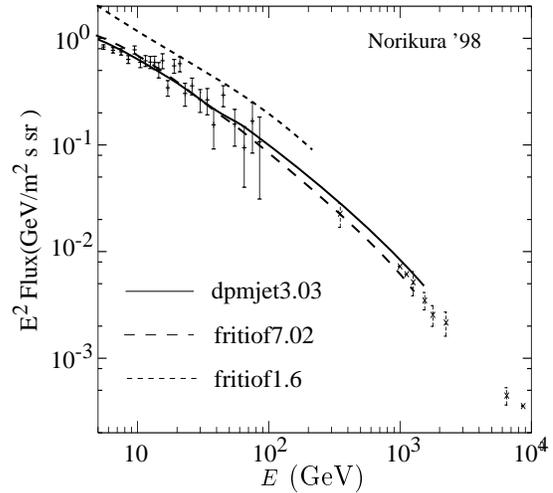} 
\caption{Gamma-ray spectrum at Mt. Norikura
 (2.77 km a.s.l). The vertical axis is Flux$\times E^2$.
Our data is at $<$ 100 GeV. 
Data above 300 GeV is from emulsion chamber
experiments. For the latter, see Sec.\ref{discuss}
}
\label{norispec}
\end{center}
\end{figure}

Figure \ref{norispec} shows the result from the observation at Mt.Norikura.   
It should be noted that the flux by fritiof1.6 becomes higher than the ones
by the other models at this altitude.

\begin{figure}[h]
\begin{center}
\includegraphics[width=7.5cm]{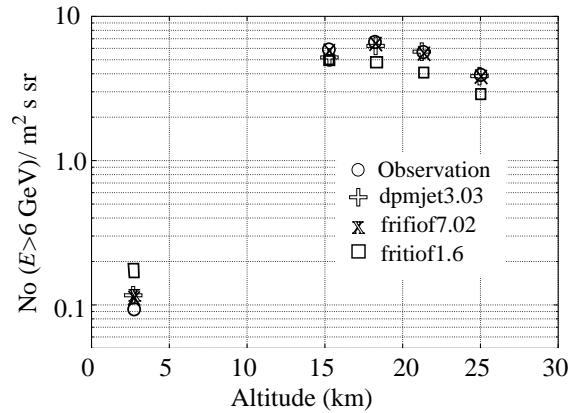} 
\caption{The altitude variation of the flux integrated
over 6 GeV.   The dpmjet3.03 and fritiof7.02 give
almost the same feature  consistent with the
observation while  the deviation of fritiof1.6 
from the data is obvious.
\label{transition}}
\end{center}
\end{figure}

From these figures, we see fritiof7.02 and dpmjet3.03 
give rapider increase
and faster attenuation of intensity  than fritiof1.6;
the tendency is very
consistent with the observed data.
The transition curve of the flux integrated over 6 GeV
shown in Fig.\ref{transition} clearly demonstrates
this feature.


\begin{center}
\includegraphics[width=6.5cm]{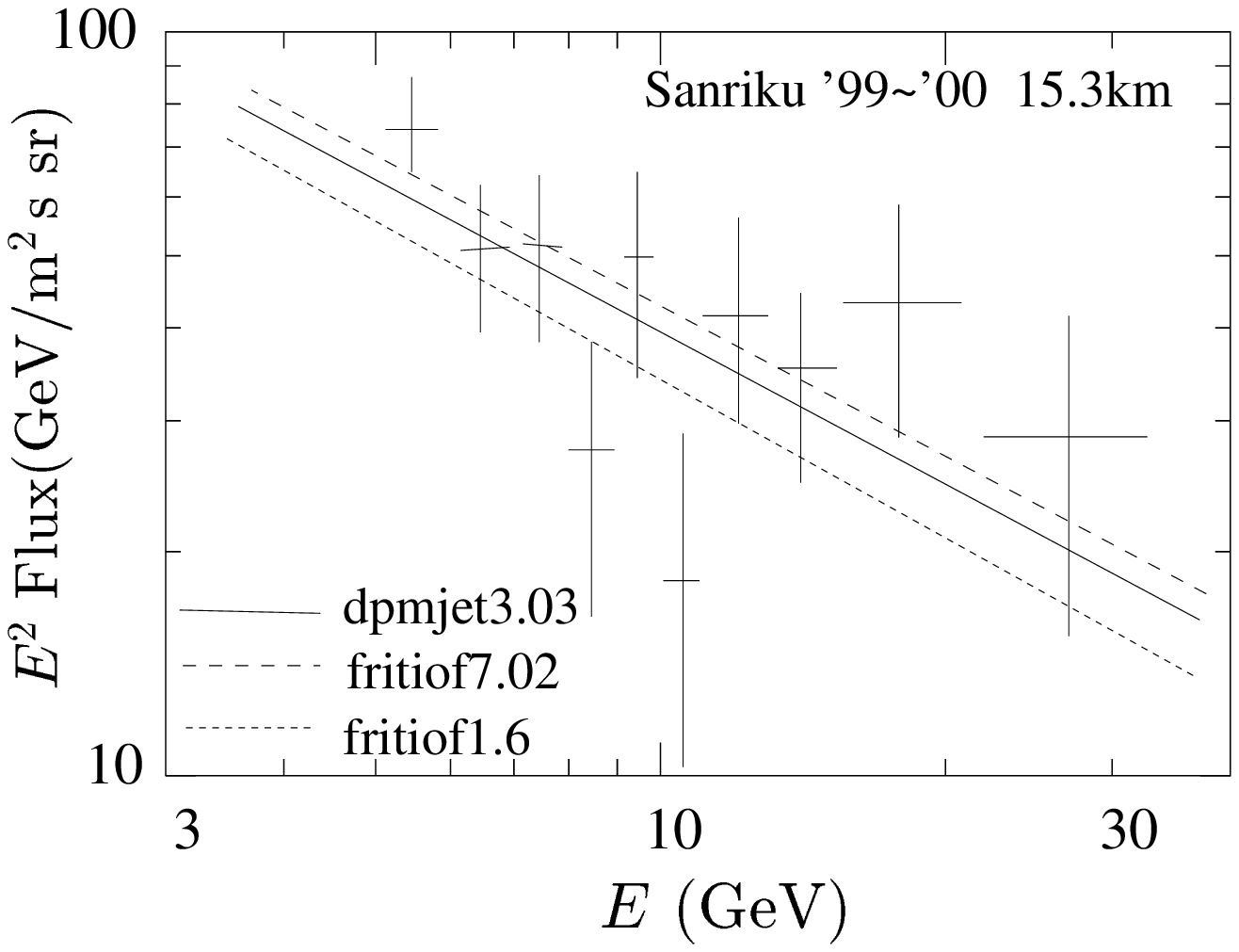} 
\includegraphics[width=6.5cm]{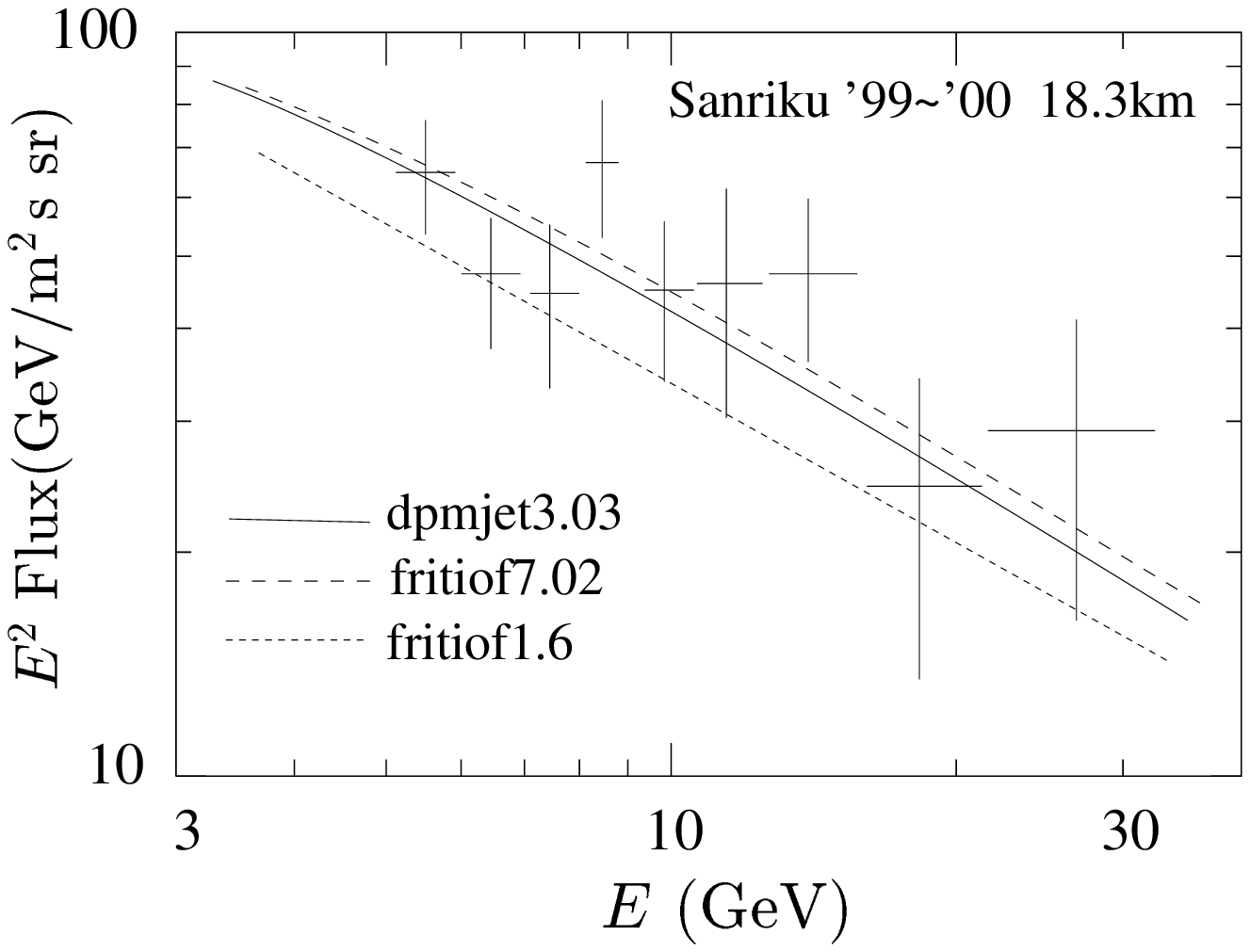} 
\includegraphics[width=6.5cm]{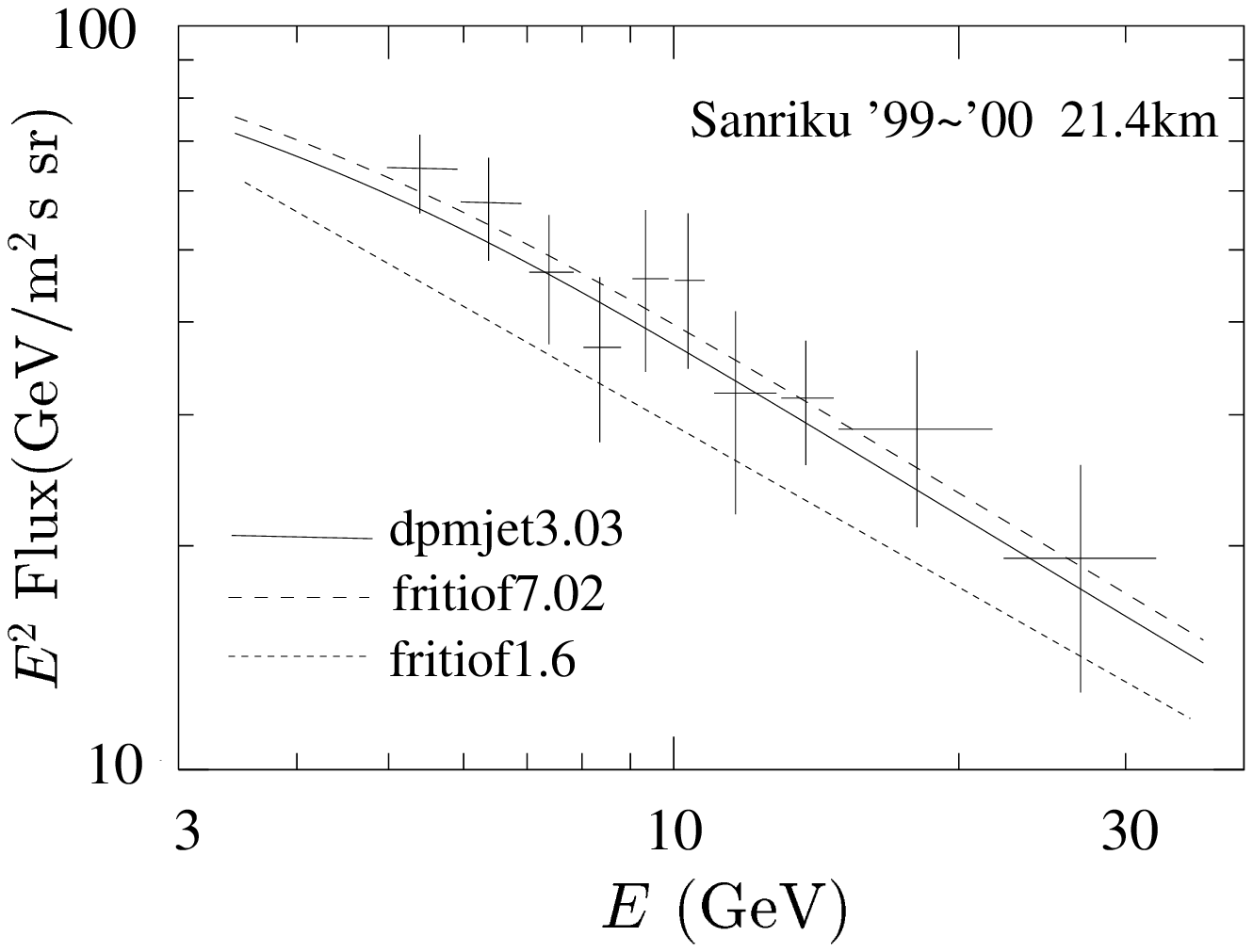} 
\includegraphics[width=6.5cm]{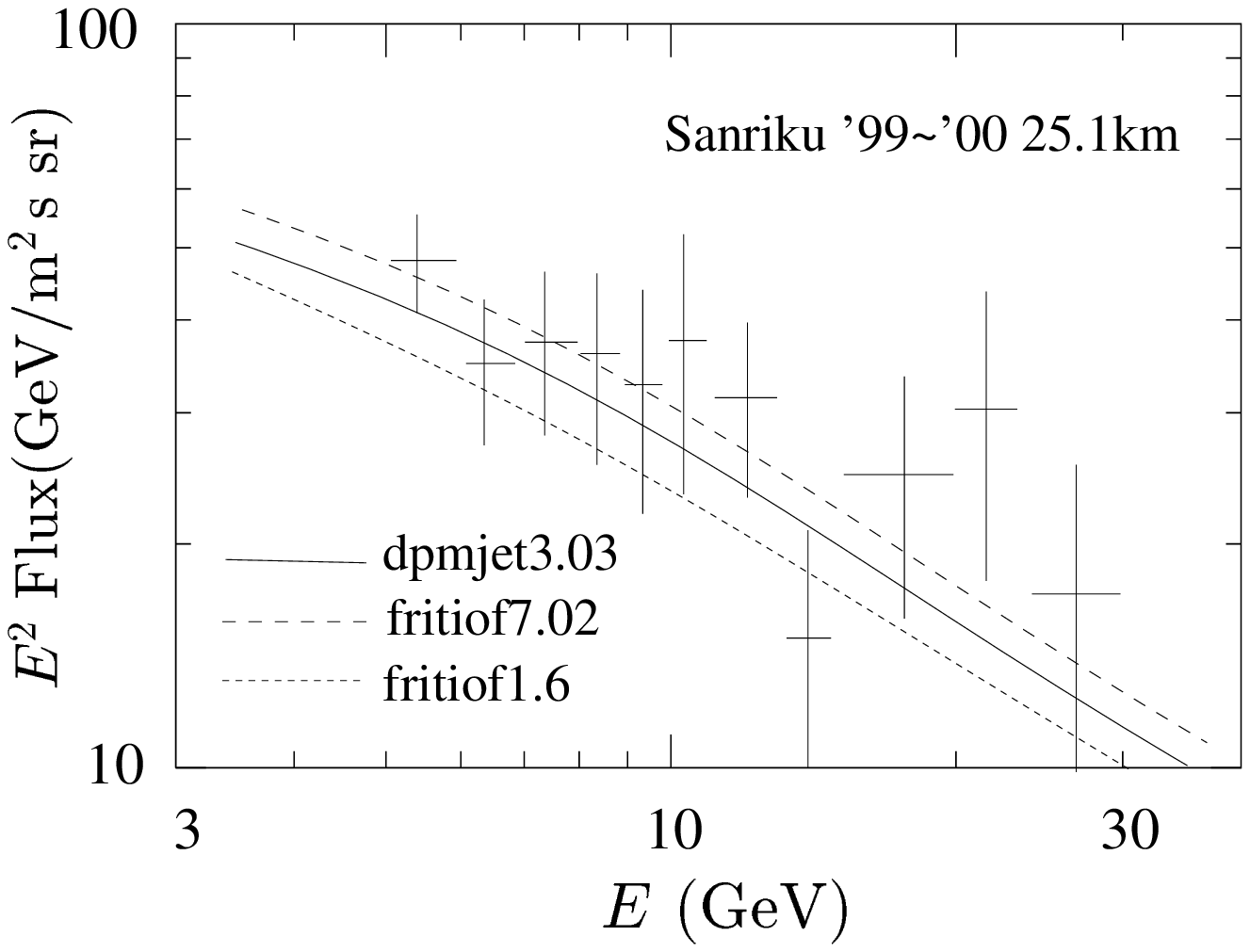} 
\end{center}
\vspace*{-1cm}

\section{Discussions\label{discuss}}

\subsection{Comparison with other data}
We found Fritiof7.02 and dpmjet3.03 give good agreement
with the observed gamma-ray data at around
10 GeV.  We briefly see whether these models
can interpret other observations.
More detailed inspection will be done elsewhere.

\begin{itemize}
\item Muon data by the BESS group at Mt.Norikura\cite{bessmuonnori}.
        
Recently, the BESS group reported detailed muon
spectrum over several hundred MeV/c.  In their paper,   calculations
by  dpmjet3.03 and fritiof1.6 are compared with the data; agreement
by dpmjet3.03 is quit good
at least above GeV where Fritiof7.02 also gives more or less the same
flux.   On the other hand, fritiof1.6 shows 
too high flux. These features are consisten with our present analysis.

\item Higher energy gamma-ray data by emulsion chamber.

        In Fig. \ref{norispec}, we inlaid an emulsion chamber
data\cite{ecc}\footnote{Electrons included in the original data 
is subtracted statistically
by use of cascade theory which is accurate at high energies.}
at Mt. Norikura. 
 Our data seems to be smoothly connected to
 their data as the two interaction models (Fritiof7.02 and dpmjet3.03)
 predict.
 Since the emulsion chamber data extends to the TeV region and the  primary
particle energy responsible for such high energy gamma-rays
is much higher than 100 GeV where we have no
accurate information comparable to the AMS and BESS data, 
it would be premature to draw a definite conclusion
on the primary and interaction model separately.
However, the fact that smooth extrapolation of
the primary spectra as shown in Table \ref{extendprim}
and the interaction model, dpmjet3.03 or fritiof7.02,
give a consistent result with the data, seems to indicate that
such combination  would provide a good estimate on other
components at $\gg$ 10 GeV.

\begin{table}[h] 
\begin{center}
\caption{Primary flux assumed in the simulation above 100 GeV/n\\
        (E in kinetic energy per nucleon (GeV),
        flux in /m$^2\cdot$s$\cdot$sr$\cdot$GeV)
\label{extendprim}
}
\begin{tabular}{|ll|ll|ll|}
\hline
\multicolumn{2}{|c|}{Proton} & \multicolumn{2}{|c|}{Helium} & \multicolumn{2}{|c|}{CNO}\\
\hline
 \ \ E &\ \   flux   &\ \    E  &\ \   flux  &\ \   E  &\ \ flux \\
\hline
92.6 & 0.593E-01  & 79.4   & 0.549E-02 & 100. & 9.0E-5 \\
108 & 0.388E-01   & 100.   & 3.0E-3 &    400.  & 1.8E-6 \\
126 & 0.276E-01   & 200.   & 5.0E-4  &  2.0E3  & 3.5E-8 \\
147 & 0.179E-01   & 400.   & 7.0E-5  &  2.0E4 &  9.3E-11 \\
171 & 0.124E-01   & 2.0E3  &  9.98E-7  &  2.0E5  & 2.3E-13   \\
200 & 0.836E-02   & 2.0E4  & 2.5E-9  &  14.0E5 & 1.3E-15  \\
1100     & 8.29E-5 & 2.0E5  &  3.97E-12  &  3.0E6 &  1.7E-16 \\
1.1E4    & 1.47E-7 & 4.0E5  &  6.1E-13  &  3.0E7 & 2.0E-19  \\
1.1E5    & 2.8E-10 & 8.0E5 &  7.0E-14  &   3.0E8   &   2.2E-22  \\
2.2E5    & 3.7E-11 & 8.0E6 &  8.7E-17  &   & \\
4.4E5    & 5.0E-12 & 8.0E8    &   5.3E-23  &   & \\
4.4E8    & 2.8E-21 & & & &\\
\hline
\end{tabular}
\end{center}
\end{table}

\end{itemize}

\subsection{The $x$-distributions}
\begin{figure}[h]
\begin{center}
\includegraphics[width=7.5cm]{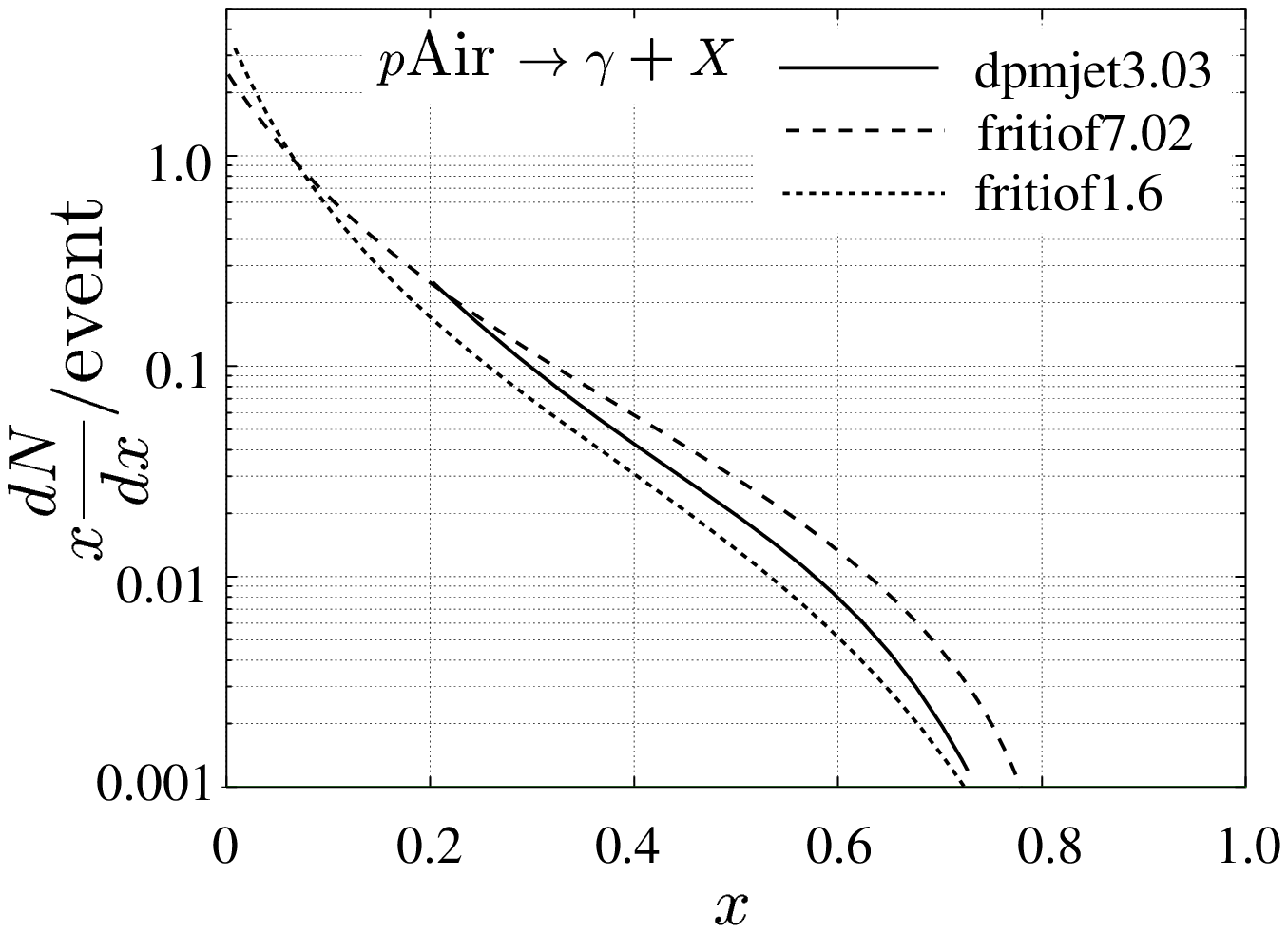} 
\includegraphics[width=7.5cm]{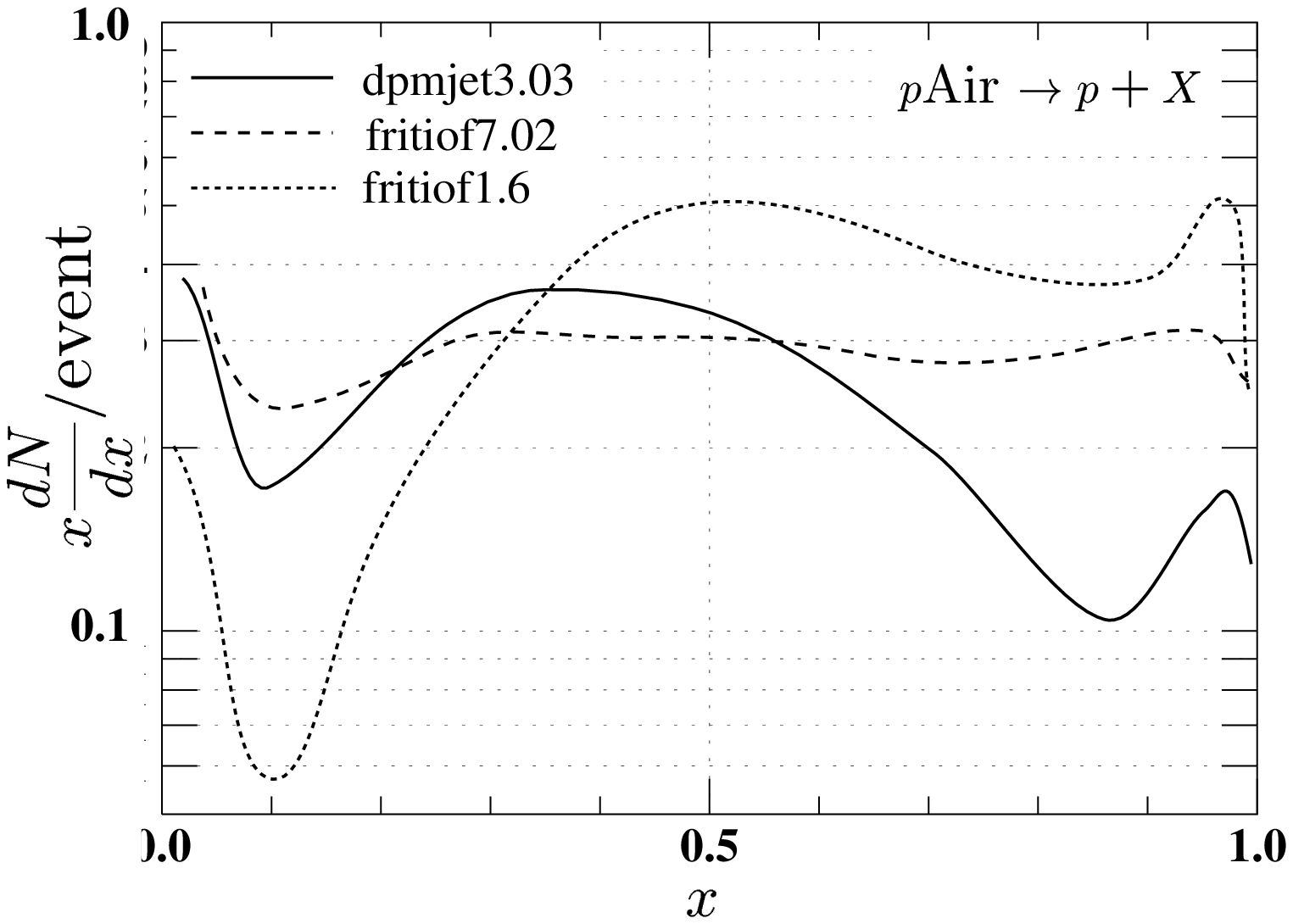} 
\caption{The $x$-distribution of photons 
from $\pi^0$ plus $\eta$ decay (upper) and
protons (lower) for $p$Air collisions
at 40 GeV.  The three model results are shown.
} 
\label{xdist}
\end{center}
\end{figure}

The two models, fritiof7.02 and dpmjet3.03, give almost the same
results in the present comparison. However, if we look into
the $x$-distribution of the particle production, we note
some difference, especially in the proton $x$-distribution.
We define the $x$ as the kinetic energy ratio of the incoming 
proton and a secondary particle in the laboratory frame.
The $x$ distribution for $p$Air collisions at  incident proton
energy of 40 GeV is presented for photons (from $\pi^0$ plus $\eta$
decay)  and 
protons in Fig.\ref{xdist}.  Difference of the  three models
seen in the photon distribution
is quite similar to the one for charged pions.
The $x$ region most effective to atmospheric gamma-ray flux is around
0.2$\sim$0.3 where the difference is not so large but
fritiof7.02 and dpmjet3.03 have
higher gamma-ray yield than fritiof1.6.

On the other hand, the proton $x$ distribution has 
larger difference among the three models (we note, however,
the difference may be exaggerated than the photon case
due to the scale difference).
It is interesting to see that,
in spite of these large differences, the final flux is not so much
different each other.  Our gamma-ray data  prefers to rather more
inelastic feature of collisions than fritiof1.6, i.e rapider
increase and faster attenuation of the flux.

We should compare the distribution with  accelerator data; however,
there is meager stuff  appropriate for our purpose. One such comparison
has been done in a  recent review paper\cite{GHreview} for $p$Air
collisions at 24 GeV/c incident momentum. The charged pion
distribution by fritiof1.6 and dpmjet3.03 well fit to some scattered data
which
prevents to tell the superiority of the two. As to the proton
distribution, among the three models,
fritiof1.6 is rather close to the data but
deviation from the data is much larger than the pion case.

The proton $x$-distribution would strongly affect the
atmospheric proton spectrum.
We calculated proton flux at Mt.Norikura to find a flux
relation such that fritiof1.6 $>$ fritiof7.02 $>$ dpmjet3.03 as expected
naturally from the $x$-distributions.  The
maximum difference is factor $\sim 2.5$ in the energy region of
0.3 to 3 GeV. The BESS group has measured the proton spectrum
at Mt. Norikura in the same energy region.
Their result expected to come soon\cite{sanukibess}
 will help select a better model for the proton $x$ distribution.

\section{summary}

\begin{itemize}
\item  We have made successful observation of atmospheric gamma-rays
at around 10 GeV at Mt.Norikura (2.77 km a.s.l) and at
balloon altitudes (15 $\sim$ 25 km). 
\item The observed gamma-ray fluxes are compared with calculations
 by three interaction models;
        it is found that fritiof1.6 employed by the HKKM calculation
        \cite{hkkm95}, which was used in 
        comparison with the Kamioka data,
        is
         not a very good model.  
\item Other two models (fritiof7.02 and dpmjet3.03) give better
         results consistent with the data, which  shows rapider
        increase and faster attenuation of the flux than fritiof1.6
        predicts.
\item Our data has complementary feature to muon data and
         will serve for checking nuclear interaction models
        used in  atmospheric neutrino calculations.
\end{itemize}

\begin{acknowledgments}
We sincerely thank the team of the Sanriku Balloon Center of the
Institute of Astronautical Science for their excellent service and
the support of the balloon flight.  We also thank the staff of
the Norikra Cosmic-Ray observatory, Univ. of Tokyo. for their
help.
We are also indebted to S.Suzuki, P.Picchi, and L. Periale for
their  spport at CERN in the beam test. For the management of
X5 beam line of SPS at CERN, we would like to thank
L. Gatignon and the tecnical staffs.  
One of the authors (K.K) thanks S. Roesler for his help in implementing
dpmjet3.03.

This work is partly supported by Grants-in Aid for Scientific
Research B (09440110), Grants-in Aid for Scientific
Research on Priority Area A (12047224) and
Grant-in Aid for Project Research of Shibaura Institute of
Technology.
\end{acknowledgments}

\bibliographystyle{apsrev}
\bibliography{betsgamma}

\end{document}